\documentclass[12pt]{article}
\usepackage{setspace}   

\usepackage{mathptmx}
\usepackage[T1]{fontenc}
\usepackage[left=1in,right=1in,top=1in,bottom=1in]{geometry} 
\usepackage{apacite} 
\usepackage{todonotes}

\usepackage{sectsty}
\usepackage{lipsum}
\allsectionsfont{\centering}

\graphicspath{{Fig/}}
\usepackage{pbox}
\usepackage{float}

\begin{document}

\title{Towards Automated Survey Variable Search and Summarization in Social Science Publications}

\author{\textbf{Yavuz Selim Kartal}\textsuperscript{1}, %
\textbf{Sotaro Takeshita}\textsuperscript{2}, %
\textbf{Tornike Tsereteli }\textsuperscript{2}, %
\textbf{Kai Eckert}\textsuperscript{4},\\
\textbf{Henning Kroll}\textsuperscript{3}, 
\textbf{Philipp Mayr}\textsuperscript{1}, 
\textbf{Simone Paolo Ponzetto}\textsuperscript{2},\\%
\textbf{Benjamin Zapilko}\textsuperscript{1}, 
\textbf{Andrea Zielinski}\textsuperscript{3} \\
   \textsuperscript{1}GESIS -- Leibniz Institute for the Social Sciences, Germany \\%
  \textsuperscript{2}Data and Web Science Group, University of Mannheim, Germany \\
  \textsuperscript{3}Fraunhofer Institute for Systems and Innovation Research ISI, Germany \\ 
  \textsuperscript{4}Web-based Information Systems and Services, Stuttgart Media University, Germany\\
  \texttt{\{tornike.tsereteli, sotaro.takeshita, ponzetto\}@uni-mannheim.de} \\ \texttt{\{yavuzselim.kartal, philipp.mayr, benjamin.zapilko\}@gesis.org} \\
  \texttt{\{henning.kroll, andrea.zielinski\}@isi.fraunhofer.de} \\
  \texttt{eckert@hdm-stuttgart.de}}

\maketitle

\begin{abstract}

Nowadays there is a growing trend in many scientific disciplines to support researchers by providing enhanced information access through linking of publications and underlying datasets, so as to support research with infrastructure to enhance reproducibility and reusability of research results.
In this research note, we present an overview of an ongoing research project, named VADIS (VAriable Detection, Interlinking and Summarization), that aims at developing technology and infrastructure for enhanced information access in the Social Sciences via search and summarization of publications on the basis of automatic identification and indexing of survey variables in text. We provide an overview of the overarching vision underlying our project, its main components, and related challenges, as well as a thorough discussion of how these are meant to address the limitations of current information access systems for publications in the Social Sciences. We show how this goal can be concretely implemented in an end-user system by presenting a search prototype, which is based on user requirements collected from qualitative interviews with empirical Social Science researchers.

\end{abstract}



\section{Introduction}

Quantitative research has been at the center of the social sciences since their inception, not least in the domains of economics and sociology. Over the years, the continuous development and refinement of theories have given rise to an equally continuous development of heuristics and the operationalisation of related, more concrete concepts. In line with the prevailing theoretical focus and the extension of quantitative research into additional domains, specific concepts of measures, or indicators, have experienced phases of inception, diffusion, and in some cases, eventual loss of prevalence. In parallel, heterogeneous sources of data have become ever more available in the past decades, thus increasing not only options for operationalisation but also those for the de facto deployment of specific indicators in diverse domains of social science research. Beyond the once paramount use of official statistics, digitalisation and, more recently, data analytics have enabled the additional use of indicators collected through surveys and/or generated from both structured or unstructured big data sources. Accordingly, the diversity of indicators used in social science studies has notably increased \cite{Diaz-Bone2020}.

From a general perspective, it is thus becoming more and more essential to better understand whether specific concepts of measurement are being used in the literature and, if so, how prevalently, so as to enable an improved information exchange between theoretical and empirical research. For those proposing theoretical advances, such information would indeed provide indication to what extent empirical research has begun to take up their new frameworks of reference. For those in charge of larger data collection or generation efforts, such information would instead provide valuable input which areas of their activities might deserve increased attention and resources and which less. Over the years, a lasting rise of certain indicators and variables has repeatedly been observed, following the proposition of seminal heuristics or guidelines subsequent to which new standards of empirical research became established. At the same time, others have remained `fashions' which after a few years of heightened research activities have faded back into oblivion. While all this is anecdotally known, a systematic review of related patterns has been so far not possible for a lack of technical options to identify and access the usage of survey variables in context, i.e., within the text of relevant scientific publications that report research using them. 

The VADIS (VAriable Detection, Interlinking and Summarization) project, on the details of which the remainder of this article will report, set out to resolve this fundamental technological challenge: our project investigates methods to deploy cutting-edge methodologies from the fields of data linking and integration, automated text understanding, and search, which we apply in order to enable improved information access of social science publications using survey variables as pivotal concept. By identifying and indexing mentions of survey variables in social science publications, we are able to provide a unified data repository in which links between heterogeneous data sources such as datasets, survey variables, and academic publications are seamlessly linked \cite{hienert2019}.

The remainder of this paper is organized as follows. We first provide an overview of our interviews with colleagues in the social sciences on the need for better (i.e., semantic) search for scholarly publications, as well as the current status of dataset search systems in Section \ref{sec:status}. We next provide a description of our research project in Section \ref{sec:vadis} and present the task of variable detection and disambiguation that is at the heart of our project in Section \ref{sec:task}. We show how information from disambiguated variable mentions in text can be used to build a search and recommendation engine for end-users in the social sciences in Section \ref{sec:search} and provide concluding remarks in Section \ref{sec:conclusion}.

\section{Searching scholarly publications in the social sciences}
\label{sec:status}

\subsection{Do social scientists need semantic search?}

In order to collect the user requirements at the heart of our project, we run a series of interviews with colleagues from the empirical social sciences (crucially including survey methodologists) at top-tier academic institutions during the second half of 2021. The main objective of the interviews was to define from an information science perspective how survey variables are used in the context of browsing, searching, and accessing academic literature. In this section, we summarize their feedback and articulated user requirement.

Just like in virtually all fields of research, in the internet era, search engines play a vital role in information access for social scientists. However, it is still difficult for the user to retrieve information from different but interconnected sources, in particular, academic literature and datasets (federated search) \cite{gregory2019,kraemer2021}. 
Additionally, retrieval is often based on proprietary software like Google Scholar, where it is difficult for the user to understand how the retrieved results are arranged based on an internal ranking algorithm (e.g., Matthew effect, see, e.g., \cite{bol18} and \cite{wang14} for a thorough analysis of this `rich gets richer' effect).

Most of all, current search engines do not support queries that explicitly relate to social science topics, concepts and relations, but only support keyword search with known limitations (i.e., vocabulary mismatch, defining complex queries, etc.). While recent advancements in automated text understanding have enabled open-domain Web search beyond keywords\footnote{\url{https://blog.google/products/search/search-language-understanding-bert}}, these models are not domain specific -- that is, they do not leverage terminology and concepts specific to the social science literature. Moreover, large-scale proprietary language models and search are known to pose non-trivial and challenging societal risks \cite{bender21}.

Crucially for our project, another major shortcoming is that concepts that are represented in heterogeneous information sources relevant for social science research -- namely, publications and survey data -- are not interlinked. Following the principles and best practices of linked data \cite{BizerHB09}, major advances in digitalised information access would come from social scientists being able not only to access research publications and data, but also publish it, by means of a platform that would allow data to be shared and reused across applications along the FAIR principles\footnote{\url{https://www.go-fair.org/fair-principles/}}, and allow reproducibility of social science studies \cite{kraemer2021}. Some key findings from \cite{kraemer2021} that are also relevant for the VADIS project, are: ``literature search is an important part of dataset search'', existing ``tools are creatively misused'', ``relevance assessment is very complex'', ``dataset search suffers from missing interlinks'' and in general ``dataset search literacy is low''.


\begin{table}[t]
    \centering
    \begin{tabular}{|c|c|c|c|}
	    \hline
	    System & Variable  & `Classic' search & Enrichment \\
	    & Retrieval & (bag-of-words) &\\
	    \hline
	    UK Data & No & Yes & No \\
	    \hline
	    ICPSR & Yes & Yes & No \\
	    \hline
	    NSD & Yes & Yes & No \\
	    \hline
	    Google Dataset & No & Yes & No \\
	    \hline
	    GESIS Search & Yes & Yes & No \\
	    \hline
    \end{tabular}
    \caption{Overview of Dataset Search Systems}
    \label{tab:1}
\end{table}

\subsection{Current dataset search systems}

In Table~\ref{tab:1} we provide a list of state-of-the-art dataset search systems in the Social Sciences  which have been mentioned and discussed by our interviewees. They include a) the UK Data Service\footnote{\url{https://ukdataservice.ac.uk/find-data/}} system which offers browsing and search facilities for studies and series. In this system the download and retrievability of subsets of data is possible, but very limited due to data owner rights (see information page\footnote{\url{https://ukdataservice.ac.uk/help/access-policy/can-i-request-specific-variables-or-subsets-of-data/}}), b) the Inter-university Consortium for Political and Social Research (ICPSR)\footnote{\url{https://www.icpsr.umich.edu/web/pages/ICPSR/ssvd/}} search system, which offers a variety of search and comparison features across variables and questions, c) the  Norwegian centre for research data (NSD)\footnote{\url{https://www.nsd.no/en/}}, which makes question texts searchable and categorizes variable types, d) Google Dataset Search\footnote{\url{https://datasetsearch.research.google.com/}}, which indexes public datasets in any domain worldwide \cite{brickley2019} and e) the GESIS Search\footnote{\url{https://search.gesis.org/}}, which integrates publications, datasets and its variables in a unified system \cite{hienert2019}. Table~\ref{tab:1} differentiates the systems regarding their support for variable retrieval (micro-level retrieval), the underlying retrieval model, and support of semantic indexing enrichment. 

What our small survey highlights is \textit{the lack of capabilities of search systems to go beyond simple keyword-based search and very little variable retrieval support, due to the lack of interlinking between publications and survey variables}. Rather than being a merely technical limitation, this precludes a variety of application scenarios that emerged from our interviews that directly impact the work of scholars in the social sciences. These include, for instance, being able to satisfy the user's information needs by ranking papers by relevance with respect to survey variables of interest, recommending related variables from a set of initial publications of interest or summarizing publications covering the same variables.

\section{VADIS project}
\label{sec:vadis}

The key vision behind VADIS\footnote{\url{https://vadis-project.github.io/}} is to allow for searching and using survey variables in context and thereby enable better information access of scholarly publications and help increase the reproducibility of research results. We will achieve this by \textit{combining text mining techniques and semantic web technologies that identify and exploit links between publications, their topics, and the specific variables that are covered in the surveys}. These semantic links in scientific texts build the basis for the development of applications to give users better access to scientific literature such as passage search, summarization \cite{takeshita22}, and information retrieval \cite{roy2022}.

\subsection{Project overview}

\begin{figure*}[t]
    \centering
    \includegraphics[width=12cm]{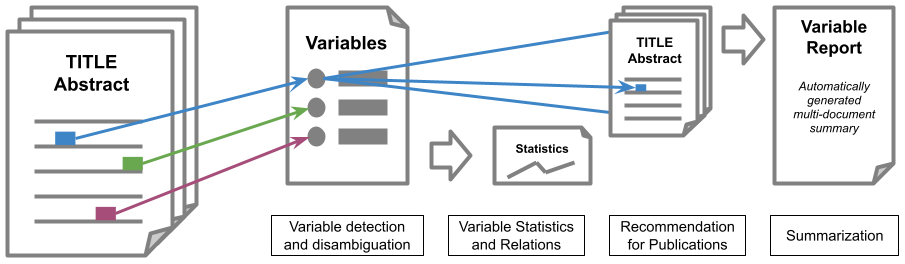}
    \caption{A top-level overview of our VADIS project: mentions of survey variables in text are automatically identified, so as to compute relevance metrics for end-user applications such as search and summarization.}
    \label{fig:process}
\end{figure*}

Figure~\ref{fig:process} provides an overview of our project. The starting point is a corpus of scientific literature (Table~\ref{tab:2}) and a vocabulary of survey variables that are used in the studies. The first step is the \textit{identification of references} to the variables in specific passages of scholarly publications. Given the survey variable mentions in text, we use this information to semantically index the documents and passages where they occur, as well as compute  \textit{variable statistics} -- e.g., estimated metrics of `importance' of a variable in a given set of papers on the basis of number of references or co-occurrence with other variable mentions in text. Since the list of publications that mention a specific variable -- or, in general, are returned as result to a user query -- can be large and can cover many topics and methods, in a final step a multi-document summary is created automatically with the goal to provide a literature overview.

\begin{table}[t]
    \centering
    \begin{tabular}{|l|r|}
	    \hline
	    \textbf{Category} & \textbf{Count}  \\
	    \hline
	    \hline
	    Publications & 117,412  \\
	    \hline
	    SSOAR Publications & 67,305  \\
	    \hline
	    Publications with Research Datasets & 13,614 \\
	    \hline
	    SSOAR Publications with Research Data & 10,106  \\
	    \hline
	    \hline
	    Research Datasets & 64,283  \\
	    \hline
	    Research Datasets with linked Variables & 547 \\
	    \hline
	    Variables & 227,246  \\
	    \hline
    \end{tabular}
    \caption{Overview of GESIS Search Corpus used in our project.}
    \label{tab:2}
\end{table}


\subsection{VADIS corpus}

In the VADIS project, we use a corpus of publications and research datasets from social science literature provided by GESIS Search \cite{hienert2019}. Table~\ref{tab:2} shows some details about the corpus (all reported numbers are from date: March 25, 2022). In total, there are 117,412 publications in this corpus from various sources. The Social Science Open Access Repository (SSOAR\footnote{\url{https://www.gesis.org/en/ssoar/home}}) is one of these sources and 57\% of the publications in GESIS Search are freely available via SSOAR as Open Access documents. These documents are the basis for our text mining activities in the project. 

Research datasets are the other essential data source for our corpus and there are 64,283 research datasets in total. Although a small part of them, 547, have digitally available survey variables, these are from the most commonly used research studies in social sciences including the German General Social Survey (ALLBUS)\footnote{\url{https://www.gesis.org/allbus/allbus}}, Eurobarometer\footnote{\url{https://europa.eu/eurobarometer/}}, the European Values Study\footnote{\url{https://europeanvaluesstudy.eu/}} and the International Social Survey Programme (ISSP)\footnote{\url{http://www.issp.org/menu-top/home/}}. 
As one random example, we mention here the ``ALLBUS/GGSS 2018'' dataset\footnote{\url{https://doi.org/10.4232/1.13250}} which is linked with over 700 distinct variables.
GESIS Search also provides the relations between the research datasets and its related publications which mention or reference this data. Regarding these relations, 13,614 of all publications are related to a research dataset and 10,106 of them are available in SSOAR.

\section{Variable detection and disambiguation}
\label{sec:task}

At the heart of our framework lies the idea that survey variables are routinely mentioned in text, and such usage in context can be automatically identified on the basis of methods from text understanding (\cite{hovy_2022}, see also \cite{grimmer_text_2013}, \cite{grimmer2022text}, and \cite{Grimmer2015WeTogether} for a critical survey of automated content analysis methods for social scientists, specifically in the context of political text analysis). Consider, for instance, the following passage from \cite{blume06}:

\textit{Of the 682 individuals in this group, 28.6\% were Catholic, 29.7\% Evangelical, 2.7\% belonged to an independent or other Christian community, and 2.5\% to a non-Christian religious community, 36.2\% of the respondents were undenominational.}

Here, we would like to acquire a model, e.g., from labeled examples using machine learning techniques, that is able to recognize that the sentence refers to the ALLBUS 2002 variable V329 RESPONDENT: RELIGIOUS DENOMINATION (``What religious confession do you belong to?'')\footnote{This variable is available in the GESIS Search via the following identifier \url{https://search.gesis.org/variables/exploredata-ZA3700\_VarV329}}.

The challenge of automatically identifying variables in a scientific publication is a particularly interesting application because the original survey variables do not occur literally in the text (as often remarked by the colleagues in the social sciences in our interviews, cf.\ Section \ref{sec:status}) and are often rephrased so that a keyword search will generally lead to miss many relevant results. Seminal work from \cite{ZielinskiM17} first introduced the task of survey variable identification in text, and proposed a methodology on the basis of textual semantic similarity. In their approach, the Variable Detection and Disambiguation task involves measuring the semantic textual similarity (STS) \cite{cer-etal-2017-semeval}, i.e., degree of content overlap, between two text snippets, namely the survey variable definitions and individual sentences in a publication. A corpus of manually labelled examples \cite{ZielinskiM18} was used to learn relevant terms, from training examples using a machine learning algorithm.

%


One of the key problems of using morpho-syntactic signals, that is, simply observing the occurrences of words in context like in \cite{ZielinskiM17}, is that these simple methods are not able to capture latent semantic factors `hidden' behind words being used to express meaning in context. As a consequence, such language representations cannot, and are not meant to capture synonymy or relevant associative relations between concepts used in the survey variable definition and its concrete usage in the scientific publication. That is, in our example, we are not able to match, for instance, `Catholic' or `Evangelical' from the publication passage to the variable definition because word-level models do not capture the knowledge that catholic and evangelical Christianity are both \textit{kinds of} religion.

To address this problem, work in Natural Language Processing (NLP) from the past decade has focused on developing a plethora of methods to learn latent semantic representations of words and larger contexts (e.g., sentences, paragraphs) from corpora. Paradigmatic examples of such representations are semantic vectors that have become \textit{de facto} ubiquitous in NLP such as Word2Vec \cite{word2vec} or BERT \cite{devlin:etal:2019}. In recent work, we accordingly evaluated up-to-date methodologies from NLP to benchmark the original results from \cite{ZielinskiM17} in light of these advances in representation learning for NLP. Our results have been collected as part of a community-wide evaluation campaign that we are currently running in the context of the third workshop on Scholarly Document Processing (SDP 2022)\footnote{\url{https://vadis-project.github.io/sv-ident-sdp2022/}}, to foster interest from the NLP community for this task.

\section{Semantic technologies for improved information access in the Social Sciences}
\label{sec:search}

Previous and current work in VADIS has focused on methods to detect and disambiguate survey variable mentions in social science publications (Section \ref{sec:task}). Within our overarching vision, this is indeed the building block on top of which more complex and end-user oriented applications can be built (cf.\ Section \ref{sec:vadis} and Figure \ref{fig:process}). That is, knowing in which papers and passages survey variables are discussed opens up a wide range of automated functionalities that are meant to enable better information access for social scientists (Section \ref{sec:status}). Specifically, with VADIS we want to enable the following end-user applications:

\begin{enumerate}
    \item[\textbf{a)}] \textbf{Search:} Identified variables can be used straightforwardly as (automatically identified) concepts to enable semantic search \cite{bast16}. As a result, we can allow users to search for publications mentioning certain variables or highly related ones.
    \item[\textbf{b)}] \textbf{Recommendation:} Knowing which survey variables are mentioned in which contexts can allow us to build, for instance, co-occurrence graphs \cite{veronis04}, which, in turn, can be used to recommend related survey variables leveraging the information that papers that used a certain variable also used other related ones.
    \item[\textbf{c)}] \textbf{Summarization:} For this, we can collect all passages from different publications discussing the \textit{same} and \textit{related} variables and compress them using techniques from NLP developed for the task of multi-document summarization \cite{Altmami20}.
\end{enumerate}

\begin{figure*}[t]
    \centering
    \includegraphics[width=16cm]{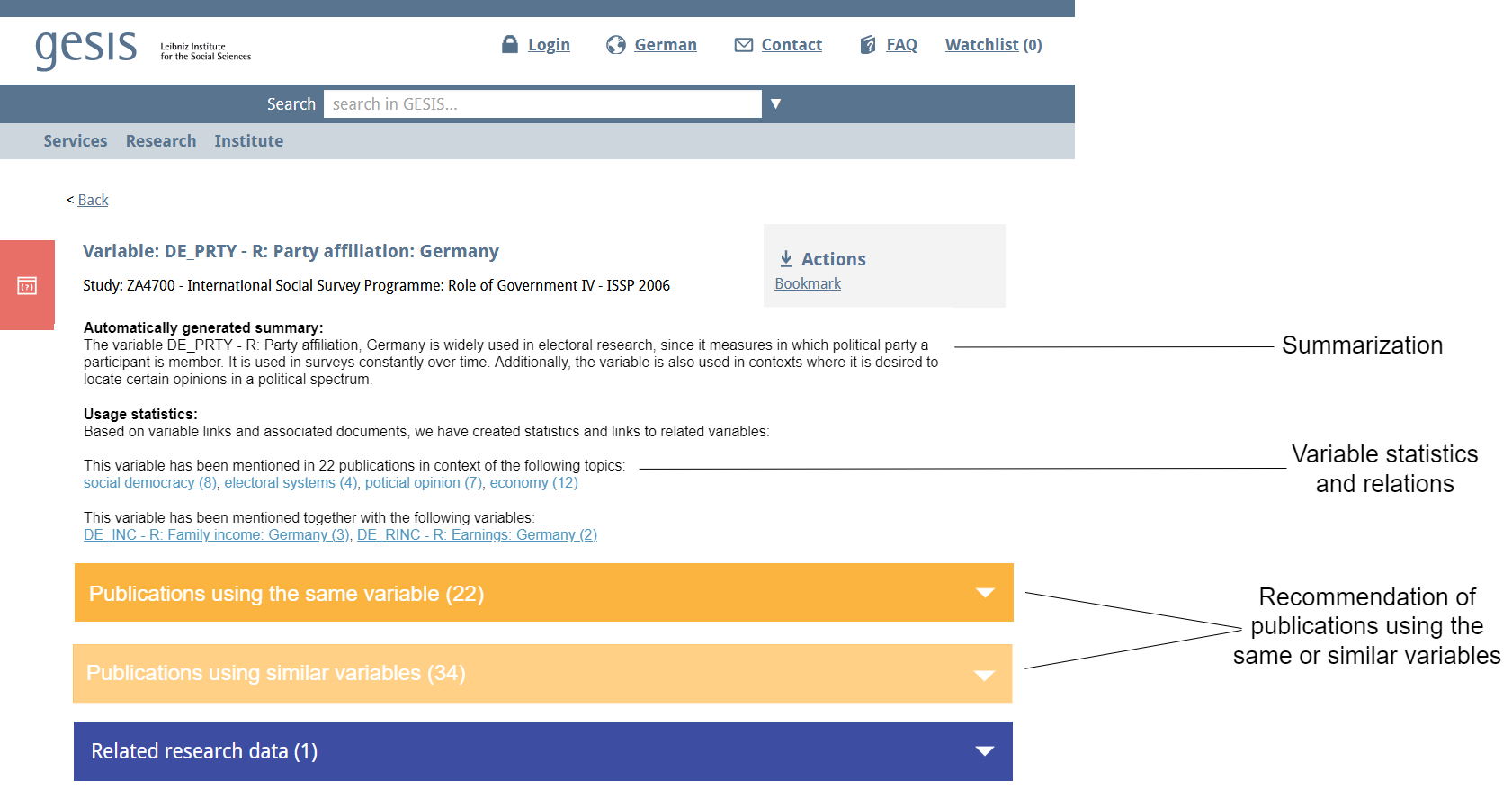}
    \caption{VADIS prototype: the system encapsulate all three main end-user applications (survey variable search, recommendation and summarization), which build upon the automatic discovery of survey variable mentions in text.}
    \label{fig:prototype}
\end{figure*}

Figure~\ref{fig:prototype} presents the prototype system that we envision as the main outcome of our project: its functionality is based on the feedback and user requirements that we collected from scholars in the social sciences during our interviews (Section \ref{sec:status}). The system encapsulates all three main end-user applications, namely survey variable search, recommendation, and summarization.  After searching and finding a certain variable which is of interest for the user, the automatically generated summary is shown which provides a textual multi-document summary on the use of the variable. Below the summary, variable statistics and relations are displayed. These statistics comprise the number of publications using the same variable, on which topics they are focusing, and other variables which have been used together with the certain variable. Below the metadata, colored bars are shown which can be expanded and contain recommendations on related publications using the same variable, publications using similar variables, and related research datasets.

\section{Conclusion}
\label{sec:conclusion}

In this paper, we presented an overview of an ongoing research project that focuses on the interlinking of survey variables and their mentions in scholarly publications. Our project has the potential to enable a new wave of semantic information access of scientific literature via semantic search, recommendation and summarization. Such technology
could have a great impact on the field in a variety of novel, exciting ways, including enabling diachronic studies of survey variable usage in publications, but also help researchers browse literature in a more serendipitous way on the basis of automated recommendation capabilities. We view these potential applications as a fundamental step toward the broader and more ambitious goal of supporting researchers in the Social Sciences by leveraging the major advances that research in data-driven text understanding (and, more generally, Artificial Intelligence) has been experiencing in the past years and decades.

\paragraph{Acknowledgement}
This work was supported by the DFG project VADIS under grant numbers: ZA 939/5-1, PO 1900/5-1, EC 477/7-1, KR 4895/3-1. 

\bibliographystyle{apacite} 
\bibliography{references}

\end{document}